\documentclass[11pt,a4paper]{article}

\usepackage[T1]{fontenc}
\usepackage[utf8]{inputenc}
\usepackage[margin=2.5cm]{geometry}
\usepackage{amsmath,amssymb,bm}
\usepackage{booktabs}
\usepackage{graphicx}
\usepackage{natbib}
\usepackage{xcolor}
\usepackage[hidelinks]{hyperref}

\newcommand{\Fn}{\mathrm{Fn}}
\newcommand{\dd}{\mathrm{d}}
\newcommand{\bx}{\bm{x}}
\newcommand{\bn}{\bm{n}}

\title{SIREN: A nonlinear Rankine-source boundary element solver for the stationary motion of
       a marine vehicle in calm water.}
\author{Daniele Peri}
\date{\small CNR-IAC -- National Research Council \\
      Istituto per le Applicazioni del Calcolo "Mauro Picone" \\
      Via dei Taurini 19, 00185 Rome, Italy \\
\texttt{d.peri@iac.cnr.it}}

\begin{document}

\maketitle

\begin{abstract}
SIREN (Ship Inviscid Resistance and Elevation eNgine)
is a Rankine-source panel solver for the steady wave resistance of
a ship, in the tradition of Dawson's method. The free-surface condition
is linearised about a base flow, and the linearisation is iterated. In the
present formulation the base flow is the solution of a Neumann problem on
the \emph{actual} boundary: sources on the entire hull, wet and dry, and
on the free surface panels, with impermeability imposed on both, so that
the free surface acts as a rigid lid at the position it currently
occupies. This replaces the classical double-body flow, whose image
about $z=0$ is meaningful only for a flat surface and a hull cut at
$z=0$, and it removes the need for desingularised free-surface panels.
In the non-linear iteration the free-surface panels are moved rigidly in
the vertical to the relaxed wave elevation at their centroid, the hull
panels are marked wet or dry against the moved surface, and the base
flow is recomputed on the moved geometry; impermeability applies to the
whole hull regardless of this marking, which is used only to select
which panels enter the force integral, and under-relaxation is confined
to the passes where the non-linear correction accumulates without being
reset. The radiation condition is imposed by a one-row upstream shift of
the collocation points, and the wave resistance is obtained by pressure
integration and, independently, by transverse wave-cut analysis. On a
Wigley-type hull, three free-surface densities and $\Fn=0.150$--$0.450$,
the non-linear iteration is stable everywhere but the lowest Froude
number on the finest grid; grid convergence, estimated by the classical
three-grid procedure, is tight for both the linear and non-linear
solutions at most speeds; and the non-linear correction raises $C_w$, the
physically expected sign, and improves agreement with towing-tank data in
the main resistance hump. Convergence to tight iterative tolerance, the
behaviour at the two highest Froude numbers on the finest non-linear
grid, and catamaran and transom-stern hulls remain open.
\end{abstract}


%=======================================================================
\section{Introduction}
%=======================================================================

The calculation of the wave-making resistance of a ship advancing
steadily in calm water is among the oldest problems of ship
hydrodynamics. The thin-ship theory of \citet{Michell1898} and the
subsequent work of \citet{Havelock1932} established the linearised
framework within which the problem was studied for most of the
twentieth century, and the parametric hull form introduced by
\citet{Wigley1934} remains the canonical test case.

The introduction of panel methods \citep{HessSmith1964} made it
possible to satisfy the hull boundary condition exactly on an arbitrary
geometry, and the decisive step for the free-surface problem was taken
by \citet{Dawson1977}, who distributed Rankine sources on both the hull
and a portion of the free surface, linearised the free-surface
condition about the double-body flow rather than about the uniform
stream, and enforced the radiation condition through an upstream-biased
finite-difference operator. Reviews of the subsequent development of
Rankine-source methods are given by \citet{Raven1996} and
\citet{Bertram2012}.

Three approximations in the classical scheme are relaxed in the present
work: the base flow about which the free-surface condition is linearised
(a double-body flow that assumes a flat surface at $z=0$), the position
of the free-surface panels, and --- least often addressed --- the
assumption that the wetted surface terminates at the still-water plane. The last
of these is of practical importance because the pressure integral that
yields wave resistance is most sensitive precisely in the bow region,
where the wave crest is highest.

Verification of a near-field pressure integration against a far-field
estimate is standard practice. The transverse wave-cut method, due to
\citet{Eggers1967} and developed by \citet{SharmaWard1968} and
\citet{Ward1964}, with the kernel formulation of
\citet{TelsteReed1994}, provides such an estimate from the same
computed field, and is implemented here.

\paragraph{Contribution.}
The contribution of the present version is a base flow that lives on
the real, current geometry. A Neumann problem is solved with sources on
the entire hull, wet and dry, and on the free-surface panels, the
latter acting as a rigid lid where they actually are, so that no image
system and no desingularisation are needed and the wetted hull may
extend above $z=0$ without overlapping its own image. The free-surface
panels are then moved rigidly to the computed wave elevation and the
hull wet/dry marking follows them, used only to select which panels
enter the force integral. Both the hull wave profile and the transverse
cuts are recomputed directly from the source strengths rather than
interpolated from panel values.

%=======================================================================
\section{Mathematical formulation}
%=======================================================================

\subsection{Governing equations}

An incompressible, inviscid fluid of density $\rho$ occupies the region
below the free surface. The flow is assumed irrotational, so that the
velocity field derives from a potential,
\begin{equation}
  \bm{u} = \nabla\Phi ,
\end{equation}
and continuity requires
\begin{equation}
  \nabla^2\Phi = 0
  \label{eq:laplace}
\end{equation}
throughout the fluid domain.

The problem is posed in a frame fixed to the ship, which advances at
constant speed $U$ along the negative $x$ axis; in this frame the flow
is steady, with $z=0$ the undisturbed free surface and $z$ positive
upwards. Lengths are made non-dimensional by the ship length $L$,
velocities by $U$, and pressures by $\rho U^2$. The Froude number is
\begin{equation}
  \Fn = \frac{U}{\sqrt{gL}} .
\end{equation}

\subsection{Boundary conditions}

On the hull surface $S_H$ the no-flux condition
\begin{equation}
  \nabla\Phi\cdot\bn = 0
  \label{eq:hull}
\end{equation}
applies, $\bn$ being the unit normal.

On the free surface $z=\eta(x,y)$ the kinematic condition requires that
the surface be a stream surface,
\begin{equation}
  \Phi_x\eta_x + \Phi_y\eta_y - \Phi_z = 0 ,
  \label{eq:kin}
\end{equation}
and the dynamic condition that the pressure there equal the atmospheric
value, which by Bernoulli's theorem gives
\begin{equation}
  \eta = -\frac{\Fn^2}{2}\left(|\nabla\Phi|^2 - 1\right) .
  \label{eq:dyn}
\end{equation}

Eliminating $\eta$ between \eqref{eq:kin} and \eqref{eq:dyn} yields the
single combined condition, to be satisfied on the \emph{a priori}
unknown surface $z=\eta$,
\begin{equation}
  \tfrac{1}{2}\,\nabla\Phi\cdot\nabla\!\left(|\nabla\Phi|^2\right)
  + \frac{1}{\Fn^2}\,\Phi_z = 0 .
  \label{eq:fs}
\end{equation}
Introducing the arc length $\ell$ along a streamline of the flow, so
that $\Phi_\ell = |\nabla\Phi| = q$, condition \eqref{eq:fs} takes the
compact form
\begin{equation}
  \Phi_\ell^2\,\Phi_{\ell\ell} + \frac{1}{\Fn^2}\,\Phi_z = 0 .
  \label{eq:fs-stream}
\end{equation}

Two further conditions close the problem: the disturbance must vanish
at infinite depth and laterally, and the radiation condition must
exclude waves ahead of the ship.

\subsection{Dawson linearisation}

Following \citet{Dawson1977}, the potential is split into a base flow
and a perturbation,
\begin{equation}
  \Phi = \Phi_0 + \varphi ,
\end{equation}
where $\Phi_0$ is the \emph{rigid-lid base flow}: the solution of
\eqref{eq:laplace} with impermeability, $(\bm{U}+\nabla\Phi_0)\cdot\bn=0$,
imposed both on the wetted hull $S_H$ and on the free-surface panels
$S_F$ in the position they currently occupy, which for the first pass
is the plane $z=0$. On a flat free surface with the hull cut at $z=0$ this
is the classical double-body problem, since the image about $z=0$ is
just a wall at $z=0$; the difference is that no image is used, so
nothing in the formulation requires the surface to be flat or the hull to
end at $z=0$. The base flow retains the leading effect of hull thickness,
which the Neumann--Kelvin linearisation about the uniform stream
\citep{BrardKelvin} does not; the latter is also available in the code
as an option.

The base flow is obtained from sources $\sigma_0$ on $S_H\cup S_F$ (the
free-surface sources of the base problem are distinct from those of the
perturbation problem below). The impermeability condition is imposed at
the centroid of each panel, with the singular self-term, on the hull and on
the lid alike; the free-surface collocation points shifted upstream for
the radiation condition play no role in the base problem, but the
base velocity is afterwards evaluated at them. Dry hull panels are
removed from the system.

Substituting into \eqref{eq:fs-stream}, retaining terms linear in
$\varphi$, and transferring the condition to $z=0$ gives
\begin{equation}
  2\,\Phi_{0\ell}\Phi_{0\ell\ell}\,\varphi_\ell
  + \Phi_{0\ell}^2\,\varphi_{\ell\ell}
  + \frac{1}{\Fn^2}\,\varphi_z
  = -\,\Phi_{0\ell}^2\Phi_{0\ell\ell}
    -\frac{1}{\Fn^2}\,\Phi_{0z} ,
  \label{eq:dawson}
\end{equation}
in which $\ell$ is now measured along the streamlines of the base flow.
On a flat lid the vertical base velocity $\Phi_{0z}$ is nonzero only
through the finite extent of the lid and the hull--lid junction, so the
last term is small at the first pass; on a moved lid it is not, and it
is retained.

\subsection{Nonlinear iteration}
\label{sec:nonlin}

The right-hand side of \eqref{eq:dawson} is the residual of the exact
condition \eqref{eq:fs-stream} evaluated on the current base flow. This
observation makes the extension to the nonlinear problem immediate: let
$\Phi_0^{(k)}$ denote the accumulated flow after $k$ passes, and solve
at each pass
\begin{equation}
  2\,\Phi^{(k)}_{0\ell}\Phi^{(k)}_{0\ell\ell}\,\varphi_\ell
  + \left(\Phi^{(k)}_{0\ell}\right)^2\varphi_{\ell\ell}
  + \frac{1}{\Fn^2}\,\varphi_z
  = -\left(\Phi^{(k)}_{0\ell}\right)^2\Phi^{(k)}_{0\ell\ell}
    -\frac{1}{\Fn^2}\,\Phi^{(k)}_{0z} ,
  \label{eq:nonlin}
\end{equation}
updating
\begin{equation}
  \Phi_0^{(k+1)} = \Phi_0^{(k)} + \omega\,\varphi^{(k)} ,
  \qquad 0<\omega\le 1 ,
\end{equation}
with $\omega$ an under-relaxation factor. At convergence
$\varphi\to 0$ and the residual vanishes, so that
\eqref{eq:fs-stream} is satisfied exactly on $z=0$. The streamline
direction is recomputed at each pass from the current flow, as are all
coefficients. Related iterative and fully nonlinear formulations are
discussed by \citet{JensenBertramSoding1989},
\citet{RavenThesis1996} and \citet{NakosSclavounos1990}.

Two levels of non-linearity are available. In the simpler one the
free-surface panels stay on $z=0$ and the iteration above only updates
the base flow: the exact condition is satisfied on the plane, not on
$z=\eta$. In the fuller one the free-surface panels are moved to the
computed wave surface, as described next, with a relaxation weight $w$
controlling how much of the newly computed elevation is adopted at
each pass, and the base flow is recomputed on the moved geometry.

\paragraph{Moving free-surface panels.}
In each of the first $n_g$ passes (the greater of the free-attitude and
moving-geometry iteration counts, both user-specified) the
elevation $\eta$ is computed from the current solution by Bernoulli's
equation at the free-surface collocation points. Since these are
shifted one row upstream, the elevation computed at the collocation point
of panel $l+n_{tra}$ is the elevation at the centroid of panel $l$
(the last row copies the previous one). The level of each panel is
updated by relaxation, $z_l\leftarrow(1-w)z_l+w\,\eta_l$, and limited
to half the draft. In the following pass the panels are moved
\emph{rigidly} in the vertical to $z_l$ --- vertices, centroid and all
auxiliary points of the local frame --- so that collocation points,
panel moments and normals (which stay $-\bm{e}_z$) are rebuilt exactly;
the wet/dry marking of the hull (\S\ref{sec:wetsurf}) is redone against the
new panel levels; and the rigid-lid base flow is solved again on the
moved geometry with the accumulated perturbation sources reset. After
pass $n_g+1$ the wet/dry marking is redone once more, against the
fully-converged wave profile rather than the still-relaxing one used
during the geometry passes (the two agree closely but not exactly);
the geometry and the base flow are then frozen, and the remaining
passes accumulate the increments as in \eqref{eq:nonlin}.

These remaining, accumulating passes are the only point in the whole
iteration where one pass's increment is not reset before the next one
is computed: every geometry pass, and the re-marking pass against the
converged wave, start the perturbation system from zero and take a
single full-strength step, which is unconditionally stable regardless
of $\omega$. Under-relaxation is therefore applied only to the
accumulating passes (adaptively, $\omega$ increased when the residual
improves and halved otherwise, never above~1 nor below the input
floor), where it removes an instability specific to the lowest Froude
number of a sweep: without it, the residual can grow across the two or
three accumulating passes and cross the divergence threshold, while
every other pass, at every Froude number, remains clean.

\subsection{Wetted surface}
\label{sec:wetsurf}

In the classical scheme the hull is truncated at $z=0$. Since the wave
elevation along the hull is of the order of the bow-wave height, and
since the pressure integral for wave resistance is a small difference
of large contributions concentrated near the bow, this truncation is
not innocuous.

The hull grid is never cut or repanelised: it is built whole, deck
included, and each panel is marked wet or dry. A panel is wet only if
\emph{all four} of its nodes lie at or below the local free-surface level,
evaluated node by node from an all-dry default. The hull therefore
never has a wet node above the adjacent free-surface panel, at the
price of a slightly under-estimated wetted surface: the error is
always by defect. In the simpler (flat) scheme the level is
$z=0$ (the classical flat cut). Once the free-surface panels are
allowed to move, the level along
the hull is a step function, equal to the level of the adjacent
free-surface panel over the longitudinal extent of that panel, so that
the marking follows the moved panels exactly and no node stands above
the panel next to it.

Impermeability, unlike the wet/dry marking above, is a purely
kinematic condition on the solid boundary: it does not depend on which
fluid, water or air, is in contact with a given panel. It is therefore
imposed on the \emph{entire} hull, wet and dry alike, at every pass;
the wet/dry marking is used only to select which panels enter the
force integral (\S\ref{sec:forceeval}), exactly as before. Earlier
versions removed dry panels from the linear system (row and column
cleared), which left a structural defect: a panel promoted from dry to
wet between one marking and the next had never actually had its row
solved while dry, so its pressure was written from a source strength
that was stale or zero, producing a spurious negative $C_p$ spike at
freshly-wetted bow panels. The spike is absent once every hull panel,
wet or dry, is a genuine unknown of the system at every pass.

Earlier versions also tested a wave profile obtained from the
sources at points off the hull, smoothed and relaxed; it produced
panels marked dry although wet and was removed together with the
associated input keyword.

\paragraph{Area-weighted partial panels.} The node-by-node marking
above is binary: a panel with some nodes above the local level and
some below is treated as entirely dry. An optional correction, the
area-weighted partial-panel force correction (default off; enabled for
every result reported in
\S\ref{sec:v8} onward), instead computes the fraction of each panel's
area that lies below the local level --- the panel is split into its
usual two triangles (as \S\ref{sec:discretisation} already does for
quadrature) and each triangle's wet area is obtained in closed form
from the signed depth at its three corners, linearly interpolated
along the edges --- and uses that fraction, rather than $0$ or $1$, to
weight the panel's contribution to the force integrals
(\S\ref{sec:forceeval}) only; it does not alter the linear system,
which already imposes impermeability on the whole hull regardless of
marking (above), nor the flow solution in any other way. This follows
the spirit, though not the exact formulas, of the partial-panel force
correction of \citet{Soding1993}. A single check at one Froude number
(Fine grid, $\Fn=0.408$) found $C_w$ change by $-1.7\%$ and $C_t$ by
$+1.3\%$ relative to the binary marking; the full three-grid,
linear-and-non-linear sweep reported below shows changes of the same
order throughout ($C_w$ typically a few percent, $C_f$ within
$1$--$4\%$, consistently positive), confirming the single check was
representative rather than a special case.

\subsection{Sinkage and trim}
\label{sec:sinkagetrim}

The vertical force and trimming moment obtained by integration of the
pressure over the wetted hull, together with the hydrostatic
restoring terms derived from the waterplane area and its second moment,
determine the equilibrium attitude. Earlier versions repanelised the
hull at each computed sinkage and trim; the present version generates
the hull panelling once, before the Froude sweep begins, and applies a
rigid rotation and translation to that single reference panelling for
every subsequent sinkage and trim. This is exact, not a small-angle
approximation: two rotations about the same axis compose by summing
angles and a vertical translation composes by summing offsets, so
applying the accumulated attitude change to the reference panelling
gives precisely the panelling a full regeneration would give. The
waterplane area and its second moment are not recomputed under this
transform and remain those of the reference attitude; every case
reported here uses a fixed attitude (sinkage and trim
not iterated), so this simplification is not exercised by the present
results.

%=======================================================================
\section{Numerical method}
%=======================================================================

\subsection{Integral representation}

The perturbation potential is represented by a distribution of Rankine
sources of unknown strength $\sigma$ on the hull and on a portion of
the free surface,
\begin{equation}
  \varphi(\bx) = \iint_{S_H \cup S_F}
       \frac{\sigma(\bm{\xi})}{|\bx-\bm{\xi}|}\,\dd S_{\bm{\xi}} ,
  \label{eq:rep}
\end{equation}
which satisfies \eqref{eq:laplace} identically. Lateral symmetry about
$y=0$ is exploited by imaging every singularity; there is no image
about $z=0$.

\subsection{Discretisation}
\label{sec:discretisation}

Both surfaces are divided into quadrilateral panels, with triangles
admitted where the hull grid degenerates. The source strength is taken
constant on each panel, and the boundary conditions are enforced at one
collocation point per panel.

Panel influence coefficients are evaluated in two regimes. For a field
point within a prescribed multiple of the panel size, the exact
integral over the plane quadrilateral or
triangle is used \citep{HessSmith1964,Newman1986}. Beyond that distance
a multipole expansion about the panel centroid is used, with the
geometric moments precomputed once and stored rather than being
re-evaluated for every source--field pair.

The sources lie \emph{on} the panels: the formulation is not
desingularised (source sheets displaced from the collocation plane, as in
\citealt{CaoSchultzBeck1991,Beck1994}). Desingularisation was used in
earlier versions and removed, together with the raising of the
free-surface panels, because it is a different formulation from the one
sought here and because its behaviour with panels moved onto the wave
surface was unsatisfactory.

\subsection{Radiation condition}

The radiation condition is imposed by staggering: the collocation point
of a free-surface panel is placed one row upstream of the panel
carrying the singularity. The resulting upstream bias plays the role of
Dawson's upwind difference operator and suppresses waves ahead of the
hull. The first row is treated by upstream extrapolation. Staggering
can optionally be extended to the transom wake strip as well as to the
main free surface; every
case reported here leaves the wake unstaggered, which
matters only for a wet transom (\S\ref{sec:transom}) and is therefore
moot for the Wigley hull, which has none. Where a transom wake is
present and left unstaggered, the upwind bias that suppresses spurious
waves ahead of the hull is absent specifically in the wake --- a
plausible, but here untested, extra source of instability for the
non-linear iteration on a transom-stern hull.

\subsection{Streamwise second derivative}

The coefficient of $\varphi_{\ell\ell}$ in \eqref{eq:dawson} requires
the second directional derivative of the influence of each panel along
the local streamline. This is evaluated analytically rather than by
finite differences over the panel network.

The same quantity is required for the base flow, $\Phi_{0\ell\ell}$,
which is obtained by summing the second-derivative influence of all
base sources $\sigma_0$ (hull and lid) and of the accumulated
perturbation sources, all with unit weight. No image is involved, so
there is no error term depending on the inclination of the streamline
with respect to $z=0$ (this was an error of first order in $\ell_z$ in
the double-body formulation once the base flow was updated).

\subsection{Linear system}
\label{sec:linsys}

Two dense, non-symmetric systems are solved at every geometry pass:
the base-flow (rigid lid) system, of order $N_H+N_F$ (every hull
panel, wet or dry, and every free-surface panel: impermeability is
imposed on the whole hull, \S\ref{sec:wetsurf}), and the perturbation
system of the same order, in which hull rows impose impermeability of
the increment and free-surface rows the linearised condition
\eqref{eq:dawson}. The default solver is an incomplete
Gaussian elimination with partial pivoting, in which coefficients below a
fraction of the pivot are skipped, followed by iterative refinement on
the resulting residual. As an option, the system is
factorised by OpenBLAS's LU routine, the solution refined in double
precision down to the same tolerance; the matrix, stored transposed, is
factorised as it is and solved with the transposed operator. The influence
matrix is allocated on the actual panel count: $N=N_H+N_F=18{,}194$ on
the Fine grid used throughout \S\ref{sec:v8} onward ($N_H=4152$,
$N_F=14042$). Filling either system is $O(N^2)$; factorising it, by
either method, is $O(N^3)$ in the worst case (the incomplete
elimination can do less work in practice, depending on how much the
drop threshold removes, but is not asymptotically better). With
the OpenBLAS option (used for every result in this paper), the
double-precision iterative refinement above took exactly two passes on
every single solve observed, base flow and perturbation alike, across
a full non-linear iteration on the Fine grid: a stable, small, and
apparently insensitive count. The
incomplete-elimination solver's own sensitivity to its drop threshold
was not tested, since the OpenBLAS option is used throughout.

%=======================================================================
\section{Transom sterns}
\label{sec:transom}
%=======================================================================

A transom that runs dry at speed leaves the hull tangentially, and the
free surface detaches from the transom edge. The condition applied on
the first rows of the wake strip is that the elevation there match the
Taylor extrapolation of the transom profile,
\begin{equation}
  \eta = z_T + \left.\frac{\partial z}{\partial x}\right|_T
              \,(x - x_T) ,
\end{equation}
linearised about the current flow in the same residual form as
\eqref{eq:nonlin}, together with a closure condition relating the
elevation at the second row to the vertical velocity at the first.
Whether the transom is wet or dry is decided by comparison of the
Froude number with a threshold $\Fn_{\mathrm{dry}}$, derived from the
transom immersion or, optionally, fixed directly by the user.

The transom face itself is not panelled, so the pressure integral over
the hull (\S\ref{sec:forceeval}) is not closed there: it is missing the
hydrostatic thrust the still-water pressure would exert on that face,
below the waterline. This is added back as a separate coefficient
$C_{yd}$, computed from the transom profile at $z=0$,
\begin{equation}
  C_{yd} = \frac{1}{\Fn^2}\int z^2 \,\dd y ,
\end{equation}
the usual hydrostatic-pressure integral over the transom width, scaled
the same way as the rest of the pressure field. In the fully-dry regime
($\Fn\ge\Fn_{\mathrm{dry}}$) no correction is added: the flow has separated
from the transom edge and there is no wetted face to close. Below the
threshold, a second integral is evaluated with the transom profile
scaled by $(\Fn/\Fn_{\mathrm{dry}})^2$, and the correction actually applied
is the \emph{difference} between the two, half-weighted, which
vanishes continuously as $\Fn\to\Fn_{\mathrm{dry}}$ from below rather than
switching on discontinuously at a hard cut-off. $C_{yd}$ enters
$C_t=C_w+C_f+C_{yd}$ with a positive sign (it adds resistance): every
case reported in this paper leaves the threshold to be determined from
the geometry, on a
transom-free Wigley hull, so $C_{yd}=0$ throughout and this mechanism
is not exercised by any result above.

%=======================================================================
\section{Force evaluation}
\label{sec:forceeval}
%=======================================================================

\subsection{Near field}

With the pressure coefficient from Bernoulli's theorem,
\begin{equation}
  C_p = 1 - |\nabla\Phi|^2 ,
\end{equation}
the wave-making resistance follows from integration over the wetted
hull,
\begin{equation}
  C_w = \frac{2}{S}\iint_{S_H}
        \left(-p\right) n_x \,\dd S ,
\end{equation}
$S$ being the total wetted surface. Vertical force and trimming moment
follow analogously.

\subsection{Frictional and total resistance}
\label{sec:friction}

Frictional resistance is obtained from two correlation lines, both
evaluated at the ship-length Reynolds number $Re_L=U_\infty L/\nu$
(freestream speed, waterline length $L$, no form factor applied to
either): the ITTC-1957 line \citep{ITTC1957},
$C_{f,\mathrm{ITTC}}=0.075/(\log_{10}Re_L-2)^2$, and, as an alternative,
a panel-by-panel flat-plate (Schlichting) integration, $C_f$, using the
converged velocity at each hull panel rather than a single ship-average
value. Total resistance is the unmodified sum of the wave, frictional
and transom hydrostatic components, $C_t=C_w+C_f+C_{yd}$; no
form-factor correction is applied to either frictional estimate. A
third, independent estimate, $C_{f,\mathrm{bl}}$, is available from the
boundary-layer march below, on a different Reynolds-number basis (local
velocity and true running length along each strip, not ship length and
freestream speed); \S\ref{sec:frictioncompare} compares all three.

\subsection{Boundary layer and transpiration velocity}
\label{sec:boundlayer}

An optional viscous correction, run once per Froude number on the
converged inviscid solution, marches an integral boundary layer along
each longitudinal panel strip using the local external velocity from
the converged potential (base flow plus non-linear correction).
Thwaites' method covers the laminar region, Michel's criterion locates
transition, and Head's entrainment method, with skin friction from the
Ludwieg--Tillmann correlation, covers the turbulent region
\citep{Schlichting1979}. From the
resulting momentum thickness and shape factor the displacement
thickness $\delta^*(s)$ is obtained, and hence an equivalent
transpiration (blowing) velocity
\begin{equation}
  V_n(s) = \frac{\dd}{\dd s}\left(U_e\,\delta^*\right) ,
\end{equation}
following \citet{Lighthill1958}. This velocity is imposed as an
additional normal-velocity boundary condition in a separate, corrective
solution of the rigid-lid Neumann problem (\S\ref{sec:nonlin}); it does
not alter the panel geometry. The resulting shift in the base-flow wave
resistance and the boundary-layer skin friction, $C_{f,\mathrm{bl}}$,
are reported alongside the inviscid and ITTC-line values.

\subsection{Comparison of frictional-resistance estimates}
\label{sec:frictioncompare}

Table~\ref{tab:cfcompare} compares the three estimates of
\S\ref{sec:friction}--\S\ref{sec:boundlayer} across the full Froude
range, Fine grid, non-linear solution. $C_f$ and $C_{f,\mathrm{ITTC}}$
agree closely, as expected of two flat-plate correlations sharing the
same Reynolds-number definition: $C_f$ is consistently $2$--$5\%$ above
$C_{f,\mathrm{ITTC}}$, and both decrease smoothly and monotonically
with $\Fn$ (rising $Re_L$). $C_{f,\mathrm{bl}}$ is a different matter:
it is $21$--$53\%$ \emph{below} $C_{f,\mathrm{ITTC}}$ at every speed,
and, unlike the other two, is not monotonic in $\Fn$ --- it rises from
$\Fn=0.150$ to a maximum near $\Fn=0.250$--$0.267$, then falls again,
tracking the local, wave-modified external velocity along each strip
rather than the ship-length Reynolds number alone. Which of the two
Reynolds-number conventions is the more physically appropriate one for
a free-surface hull is not settled here; the comparison is offered as
a diagnostic; $C_{f,\mathrm{bl}}$ is not used in $C_t$ anywhere above.

\begin{table}[htbp]
\centering
\caption{Frictional resistance coefficient $C_f\times10^3$ by three
methods, Fine grid, non-linear solution, area-weighted partial-panel
correction enabled.
$C_{f,\mathrm{ITTC}}$: ITTC-1957 line; $C_f$: panel-by-panel
flat-plate integration, ship-length $Re_L$; $C_{f,\mathrm{bl}}$:
boundary-layer march, local velocity and running length.}
\label{tab:cfcompare}
\begin{tabular}{lrrrr}
\toprule
$\Fn$ & $C_{f,\mathrm{ITTC}}$ & $C_f$ & $C_{f,\mathrm{bl}}$ & $C_{f,\mathrm{bl}}/C_{f,\mathrm{ITTC}}$ \\
\midrule
0.150 & 3.674 & 3.728 & 1.813 & 0.49 \\
0.180 & 3.549 & 3.618 & 2.144 & 0.60 \\
0.211 & 3.444 & 3.529 & 2.297 & 0.67 \\
0.250 & 3.338 & 3.441 & 2.638 & 0.79 \\
0.267 & 3.298 & 3.402 & 2.560 & 0.78 \\
0.289 & 3.251 & 3.359 & 2.441 & 0.75 \\
0.316 & 3.199 & 3.329 & 2.281 & 0.71 \\
0.354 & 3.135 & 3.259 & 2.024 & 0.65 \\
0.408 & 3.058 & 3.188 & 1.798 & 0.59 \\
0.450 & 3.006 & 3.169 & 1.598 & 0.53 \\
\bottomrule
\end{tabular}
\end{table}

\paragraph{Form factor (Prohaska).} A form-factor correction
$(1+k)C_{f,\mathrm{ITTC}}$, with $(1+k)$ fitted by the ITTC-1978
Prohaska procedure (linear regression of $C_t/C_{f,\mathrm{ITTC}}$
against $\Fn^4/C_{f,\mathrm{ITTC}}$ at Froude numbers low enough that
wave resistance is negligible, extrapolated to $\Fn=0$), was considered
as a refinement of the frictional estimate used to extract
$C_{w,\mathrm{exp}}$ below. It is not usable with the present
experimental data set: only two Froude numbers, $\Fn=0.150$ and
$0.180$, lie below the resistance hump, and the two-point regression
they define gives $(1+k)\approx0.97$ --- less than unity, i.e.\ a
\emph{negative} viscous form correction, which is unphysical and is
itself evidence that two points this close to the hump are not in the
regime the method assumes. Prohaska's procedure needs several points
spread over a genuinely wave-resistance-free range (typically
$\Fn\lesssim0.1$--$0.15$ for a hull of this length) to fit a reliable
slope; this data set does not extend low enough. $C_{f,\mathrm{ITTC}}$
without a form factor is used throughout, as already stated in
\S\ref{sec:friction}.

\subsection{Far field: transverse wave-cut analysis}
\label{sec:wavecut}

An independent estimate of the wave resistance is obtained from the
free-wave spectrum. Scaling lengths by $U^2/g$, so that the fundamental
wave number is unity, a free wave of transverse wave number $u$ has
longitudinal wave number $w$ satisfying
\begin{equation}
  w^2 = \sqrt{w^2+u^2}
  \qquad\Longrightarrow\qquad
  w^2 = \tfrac12\left(1+\sqrt{1+4u^2}\right) .
  \label{eq:disp}
\end{equation}
On a transverse cut at $x=x_t$, the cosine transforms of the elevation
and of its streamwise derivative,
\begin{equation}
  C(u)  = \int \eta\,\cos(uy)\,\dd y ,
  \qquad
  C_x(u) = \int \eta_x\,\cos(uy)\,\dd y ,
\end{equation}
combine into the free-wave amplitude
\begin{equation}
  H(u) = 2\left[C(u) + \mathrm{i}\,\frac{C_x(u)}{w}\right]
         \mathrm{e}^{\mathrm{i}wx_t} ,
\end{equation}
the factor of two accounting for the half cut of a laterally symmetric
field, and the resistance follows as
\begin{equation}
  R_w = \frac{1}{4\pi}\int_0^{\infty}
        \frac{\sqrt{1+4u^2}}{1+\sqrt{1+4u^2}}\,\left|H(u)\right|^2
        \,\dd u .
  \label{eq:rwcut}
\end{equation}
The kernel is that of \citet{TelsteReed1994}; the equivalent
formulation in terms of the wave angle is given by
\citet{SharmaWard1968}.

Because the source strengths are known, the cut is not sampled from
panel values but recomputed: the velocity, and hence the elevation
through \eqref{eq:dyn}, is evaluated directly from \eqref{eq:rep} at
arbitrary points, on a uniform transverse grid independent of the panel
layout, with $\eta_x$ obtained by central differences over a step
chosen from the wavelength. The same device supplies the wave profile
along the hull used for the wetted-surface cut.

\paragraph{Resolution limit.}
The integration in \eqref{eq:rwcut} must be truncated. Although the cut
may be sampled arbitrarily finely, it carries no information finer than
the panels that generated the solution; integrating to the sampling
Nyquist limit admits panel-scale structure of the near field, which the
kernel weights as heavily as the physical spectrum since
$\sqrt{1+4u^2}/(1+\sqrt{1+4u^2})\to 1$ as $u\to\infty$. The truncation
is therefore set by the transverse panel size of the free-surface grid.

\paragraph{Consistency check.}
The free-wave spectrum is independent of the cut abscissa. The variation
of $C_w$ across cuts is therefore a direct measure of the reliability
of the estimate, and is reported as such.

\subsection{Lifting correction}
\label{sec:lifting}

An optional approximate lifting-line correction, intended for planing
or semi-planing hulls where dynamic lift is a significant fraction of
the vertical force, adds an induced-drag and lift contribution derived
from an equivalent finite wing of aspect ratio $A\!R=B^2/S$ ($B$ beam,
$S$ wetted surface), with lift slope
\begin{equation}
  \frac{\dd C_L}{\dd\alpha} = 2\pi\,\frac{A\!R}{A\!R+2}
\end{equation}
and induced drag $C_i = C_L^2/(\pi A\!R)$. The user specifies the
section lift coefficient $C_{L0}$ directly, and selects how the
effective angle of
attack $\alpha$ is obtained, among: the apparent bow/stern angle,
$\arctan(-z_{\min}/(x_{\max}-x_{\min}))$; the angle implied by the
longitudinal distribution of the submerged-volume centroid depth;
the current trim angle; a fixed unit angle; or the correction switched
off entirely. The resulting lift and moment, $C_L=2\pi C_{L0}\alpha\,
A\!R/(A\!R+2)$ and $C_M=-c_M\,C_L$ (with $c_M$ a user-specified
proportionality constant), are added to the vertical
force and trimming moment as $C_L\,S/2$ and $C_M\,S/2$, and the induced
drag $C_i\,S/2$ is added to the wave resistance. The correction is
purely geometric (through $A\!R$ and, for the first two angle choices, the
hull shape) or attitude-dependent (for the trim-angle choice); it does not
depend on the solved potential.

%=======================================================================
\section{Implementation}
%=======================================================================

\subsection{Architecture}

SIREN is a single Fortran executable of approximately $18$ thousand
lines. Hull and free-surface grid generation are
subroutine libraries linked into the solver rather than separate
programs, and all data passes through memory: there are no intermediate
files. Input files are read once at start-up and retained, so that the
hull may be regenerated at every pass of the geometry phase without
disk access.

The solver is organised around the stages described above: assembly and
solution of the rigid-lid base flow, assembly and solution of the
perturbation system (including the streamline second derivatives of
\S\ref{sec:linsys}), the rigid vertical move of the free-surface panels
and the associated wet/dry marking, the choice between the two linear
solvers, and the near- and far-field force evaluation. The input file
selects the formulation through only a handful of settings: the number
of non-linear passes, the under-relaxation factor, the number of
geometry-tracking passes, the number of free-attitude iterations, the
free-surface-panel relaxation weight, and the choice of linear solver.

This matters beyond convenience. The historical file-based exchange
quantised the geometry to the precision of its output formats at every
regeneration, which is acceptable for a single computation but not for
an iterative scheme in which the error accumulates.

\subsection{Memory}

Array dimensions are determined at run time. A panel buffer is
allocated, the geometry generated, and the buffer doubled on overflow;
both generators test their limits before writing. The influence matrix,
whose size grows as the square of the panel count and dominates the
requirement, is allocated on the actual count.

\subsection{Parallelisation}

The assembly of the influence matrix and the field evaluations are
parallelised with OpenMP over the panel loop, and the OpenBLAS solver
(\S\ref{sec:linsys}) parallelises the factorisation and refinement
internally. On a single Froude number, Coarse grid, this machine (16
cores), wall-clock time dropped from $379.5$~s at one thread to
$50.5$~s at sixteen: a $7.5\times$ speed-up, about $47\%$ of the ideal
$16\times$, consistent with a non-trivial serial fraction (grid
generation, input/output, and any part of the iteration not explicitly
threaded) rather than a parallelisation defect specific to the panel
loop. A phase-by-phase breakdown of run time between assembly,
solution and field evaluation was not instrumented and is not reported.

\subsection{Output}

Results are reported both as non-dimensional coefficients and as
dimensional forces derived from the ship length. Three-dimensional
fields are written as VTK unstructured grids, separately for the hull
--- carrying pressure --- and for the free surface --- carrying
elevation --- so that each may be visualised on its own scale, with the
hull in its computed attitude.

\subsection{Additional capabilities}
\label{sec:capabilities}

The following are implemented and selectable through the input files
but are not exercised by the results reported below; they are listed
here, without evaluation, so that later work can refer to this paper
for their description.

\begin{itemize}
\item \textbf{Twin-hull (catamaran/SWATH) geometry}: an option in the
  grid generator produces two mirrored hull domains offset by a
  specified half-beam, with the
  free-surface generation adjusted for the gap between them.
\item \textbf{Transom dry threshold}: the Froude-number threshold above
  which the transom is treated as dry (\S\ref{sec:transom}) can either
  be fixed directly by the user or
  left to be determined from the transom immersion.
\item \textbf{Exact wave-profile waterline row}: an alternative to the
  node-by-node wet/dry marking of \S\ref{sec:wetsurf}, which resamples
  each hull section so that one grid node falls exactly on the computed
  wave elevation, giving an exact rather than a stepped wet/dry
  boundary.
\item \textbf{Wave-cut resolution and upstream anchoring}: the number of
  transverse cuts and of points per cut used by the far-field analysis
  of \S\ref{sec:wavecut} are set independently of the panel grid;
  the most upstream free-surface rows can optionally
  be held fixed at $z=0$ instead of following the local solution,
  to keep the upstream radiation condition
  anchored on an undisturbed row.
\item \textbf{Free-surface/hull lateral offset}: the free-surface mesh
  can be offset laterally from the hull and centreplane by a fraction
  of the local panel width, to
  avoid a zero-gap contact; the default is no offset.
\item \textbf{Geometric lifting-surface panels}: extra panel rows can be
  extruded at a specified vertical offset from selected hull sections,
  for appendages such as foils or spray rails;
  this is a purely geometric addition to the panelled surface, distinct
  from the hydrodynamic lifting-line correction of \S\ref{sec:lifting}.
\item \textbf{Geometry-control output}: an optional output mode writes
  intermediate grid-generation
  files for inspection independent of the solver run.
\end{itemize}

%=======================================================================
\section{Verification and validation}
%=======================================================================

\subsection{Results with the present formulation}
\label{sec:v8}

\paragraph{Set-up.}
Three free-surface densities were used, in a constant refinement ratio
$r=1.5$: 24, 36 and 54 panels along the hull length (Coarse, Medium,
Fine), with the upstream fetch and the longitudinal panel count
upstream of the hull scaled by the same $r$ on all three, so that
free-surface density is the only quantity being refined; the hull grid
itself (the $100\times30$ node file, generated once, \S\ref{sec:wetsurf})
is identical on all three. Each grid was run twice, with the OpenBLAS
solver: linearised (the classical
scheme, no panel motion) and non-linear (ten passes, moving
free-surface panels with under-relaxation, \S\ref{sec:nonlin}), fixed
attitude throughout, and with the area-weighted partial-panel force
correction
enabled (\S\ref{sec:wetsurf}): a single-case check found it changes $C_w$ by a
few percent without altering the flow solution, and it is included
below as the more accurate of the two force-integration options.
Coarse and Medium omit $\Fn=0.150$ (Coarse also $0.180$): at these
speeds the free-surface wavelength is not resolved and the
acute-divergence guard rejects the run before it reaches the fixed
point. The eight Froude numbers common to all three grids,
$\Fn=0.211$--$0.450$, are used for the grid-convergence study below;
Fine alone was also run at $0.150$ and $0.180$.

\paragraph{Stability.}
Every one of the $3\times10=30$ runs (three grids, ten Froude numbers
each, non-linear) reached its final iteration without triggering the
divergence guard, except Fine at $\Fn=0.150$, its own lowest speed,
which the guard rejects after two accumulating passes; the corresponding
Coarse and Medium runs at that speed are the ones omitted above. This
is a qualitative change from every non-linear configuration tried
earlier in this project (\S\ref{sec:nonlin-eff}), all of which diverged
somewhere in $\Fn=0.211$--$0.450$ on at least one of the three grids.

\begin{table}[ht]
\centering
\caption{Coarse, pressure-integrated $C_w\times10^3$: linearised and
non-linear, OpenBLAS solver, present formulation, area-weighted
partial-panel correction
enabled. Ratio: non-linear/linear.}
\label{tab:v8sweep}
\begin{tabular}{crrr}
\toprule
$\Fn$ & linear & non-linear & ratio \\
\midrule
0.211 & 0.734 & 0.802 & 1.09 \\
0.250 & 0.989 & 0.995 & 1.01 \\
0.267 & 0.853 & 0.824 & 0.97 \\
0.289 & 1.350 & 1.326 & 0.98 \\
0.316 & 1.647 & 1.729 & 1.05 \\
0.354 & 1.271 & 1.377 & 1.08 \\
0.408 & 2.325 & 2.320 & 1.00 \\
0.450 & 3.391 & 3.314 & 0.98 \\
\bottomrule
\end{tabular}
\end{table}

With the area-weighted partial-panel correction enabled, the non-linear correction on the Coarse
grid is close to neutral --- $\pm9\%$, split roughly evenly between
raising and lowering $C_w$ --- rather than uniformly raising it, as it
still does with the plain binary wet/dry marking (not shown): the two
force-integration options do not just rescale the same curve, they
change which sign the non-linear correction takes at several Froude
numbers. Neither result is obviously wrong on its own terms; which one
tracks physical reality better is not established here. Finer grids
are examined next.

\subsection{Grid convergence}
\label{sec:gridconv}

Grid uncertainty was estimated with the classical three-grid
Richardson procedure \citep{Roache1994,Celik2008}: apparent order
$p=\ln(\varepsilon_{32}/\varepsilon_{21})/\ln r$, Richardson-extrapolated
estimate $f_\infty = f_1 + (f_1-f_2)/(r^p-1)$, and grid convergence
index $\mathrm{GCI}=F_s\,|\varepsilon/f|/|r^p-1|$ with the recommended
three-grid safety factor $F_s=1.25$, subscripts $1,2,3$ denoting
Fine, Medium and Coarse. The procedure requires $C_w$ to vary
monotonically with grid density; where it does not (below), no $p$,
$f_\infty$ or GCI is reported.

\paragraph{Linear.} Table~\ref{tab:gridconv} gives $C_w$ on the three
densities for the linearised scheme, together
with the apparent order, Richardson value and GCI where defined, and
the thin-ship value of \citet{Michell1898} where available. Six of the
eight Froude numbers are monotonic; $\Fn=0.250$ and $0.267$ are not,
the three grids agreeing to within $1$--$2\%$ of $C_w$ itself; where
GCI is defined it is small, $0.1$--$4\%$, an order of magnitude tighter
than under the formulation of an earlier version of this project (not
reported here, superseded), consistent with the removal of the dry-panel
exclusion from the linear system (\S\ref{sec:wetsurf}), which affects
the linear solution exactly as it does the non-linear one.

\begin{table}[htbp]
\centering
\caption{Grid convergence of the linear wave resistance coefficient
$C_w\times10^3$, Wigley hull, linearised scheme, area-weighted
partial-panel correction
enabled. Coarse/Medium/Fine: 24/36/54 free-surface panels along the
hull length ($r=1.5$). $p$: apparent order; $f_\infty$: Richardson
extrapolation from Medium/Fine; $\mathrm{GCI}_{21}$, $\mathrm{GCI}_{32}$:
grid convergence index, fine and coarse pair ($F_s=1.25$); asym.:
asymptotic ratio $\mathrm{GCI}_{32}/(r^p\mathrm{GCI}_{21})$. Michell:
thin-ship value of \citet{Michell1898}. ``n.m.'': not
monotonic, no GCI defined.}
\label{tab:gridconv}
\begin{tabular}{lrrrrrrrrr}
\toprule
$\Fn$ & Michell & Coarse & Medium & Fine & $p$ & $f_\infty$ &
  $\mathrm{GCI}_{21}$ (\%) & $\mathrm{GCI}_{32}$ (\%) & asym. \\
\midrule
0.211 & --    & 0.734 & 0.696 & 0.684 & 2.84 & 0.678 & 1.0 & 3.1 & 0.98 \\
0.250 & 1.064 & 0.989 & 1.005 & 0.995 & \multicolumn{5}{c}{n.m.} \\
0.267 & 0.976 & 0.853 & 0.837 & 0.855 & \multicolumn{5}{c}{n.m.} \\
0.289 & 1.987 & 1.350 & 1.415 & 1.435 & 2.90 & 1.444 & 0.8 & 2.5 & 1.01 \\
0.316 & 1.831 & 1.647 & 1.654 & 1.657 & 2.90 & 1.658 & 0.1 & 0.3 & 1.00 \\
0.354 & 1.284 & 1.271 & 1.273 & 1.294 & (low $p$) & 1.271 & 2.2 & 0.2 & 1.02 \\
0.408 & 3.030 & 2.325 & 2.358 & 2.381 & 0.87 & 2.436 & 2.9 & 4.1 & 1.01 \\
0.450 & 4.154 & 3.391 & 3.376 & 3.367 & 1.32 & 3.354 & 0.5 & 0.8 & 1.00 \\
\bottomrule
\end{tabular}
\end{table}

\paragraph{Non-linear.} The three grids were rerun with the coupled
scheme of \S\ref{sec:nonlin} active (ten non-linear and geometry-tracking
passes,
an under-relaxation factor of $0.3$, tail-scoped under-relaxation as described there).
Table~\ref{tab:gridconv-nl} gives the results, same layout as
Table~\ref{tab:gridconv}. Two of the eight Froude numbers,
$\Fn=0.211$ and $0.289$, are not monotonic, at the same level of
agreement as the linear case ($C_w$ within a few percent of itself
across the three grids); $\Fn=0.250$, $0.267$ and $0.316$ are monotonic
with a well-defined, small GCI ($0.3$--$8\%$); $\Fn=0.354$--$0.450$ are
monotonic but with a low or negative apparent order, so the Richardson
value is not meaningful there (marked ``low $p$'') --- at all three
Fine departs from Coarse/Medium in a way the two-grid extrapolation
cannot resolve, most severely at $\Fn=0.408$ (Fine $24\%$ below
Coarse). This pattern (three clean mid-range Froude numbers, trouble at
the low and high ends) is different in detail from the binary-marking
result, but the qualitative picture --- most of the range well behaved,
the top of the range grid-inconsistent on Fine --- is the same.

\begin{table}[htbp]
\centering
\caption{Grid convergence of $C_w\times10^3$ under the non-linear
iteration (ten non-linear and geometry-tracking passes, an
under-relaxation factor of $0.3$),
area-weighted partial-panel correction enabled, same grids and layout as Table~\ref{tab:gridconv}.
``n.m.'': not monotonic; ``low $p$'': $p<0.5$, Richardson value not
meaningful, Fine alone reported.}
\label{tab:gridconv-nl}
\begin{tabular}{lrrrrrrrr}
\toprule
$\Fn$ & Coarse & Medium & Fine & $p$ & $f_\infty$ &
  $\mathrm{GCI}_{21}$ (\%) & $\mathrm{GCI}_{32}$ (\%) & asym. \\
\midrule
0.211 & 0.802 & 0.685 & 0.710 & \multicolumn{5}{c}{n.m.} \\
0.250 & 0.995 & 1.059 & 1.092 & 1.62 & 1.128 & 4.1 & 8.1 & 1.03 \\
0.267 & 0.824 & 0.867 & 0.881 & 2.86 & 0.887 & 0.9 & 2.9 & 1.02 \\
0.289 & 1.326 & 1.390 & 1.358 & \multicolumn{5}{c}{n.m.} \\
0.316 & 1.729 & 1.784 & 1.798 & 3.48 & 1.802 & 0.3 & 1.2 & 1.01 \\
0.354 & 1.377 & 1.412 & 1.453 & \multicolumn{5}{c}{low $p$} \\
0.408 & 2.320 & 2.210 & 1.791 & \multicolumn{5}{c}{low $p$} \\
0.450 & 3.314 & 3.123 & 2.841 & \multicolumn{5}{c}{low $p$} \\
\bottomrule
\end{tabular}
\end{table}

\subsection{Comparison with experiment and Michell}
\label{sec:linref}

\begin{figure}[htbp]
\centering
\includegraphics[width=0.95\textwidth]{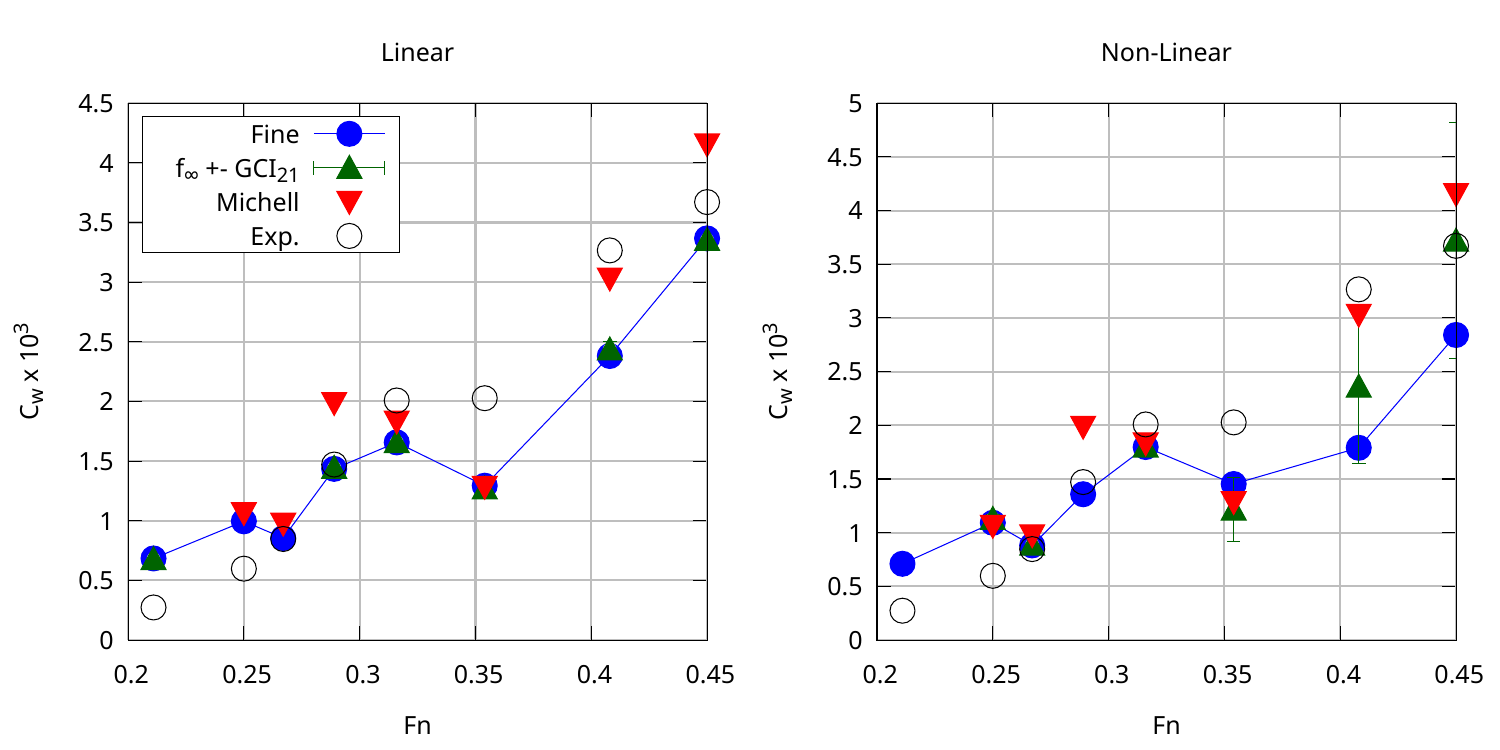}
\caption{Fine-grid $C_w$, asymptotic value $f_\infty$ with its
$\mathrm{GCI}_{21}$ band where defined, Michell and experiment; linear
(left) and non-linear (right).}
\label{fig:cwconv}
\end{figure}

Table~\ref{tab:exp} compares $C_t=C_w+C_{f,\mathrm{ITTC}}$, computed
from Fine and, where Table~\ref{tab:gridconv}/\ref{tab:gridconv-nl}
define it, from $f_\infty$, with the Wigley-hull towing-tank data of
\citet{Kajitani1983} (Ship Research Institute, Tokyo facility, $L=4$~m,
$C_{w,\mathrm{exp}}=C_{t,\mathrm{exp}}-C_{f,\mathrm{ITTC}}$, a small
difference of large numbers at low Froude number). Figure~\ref{fig:cwconv}
plots the same comparison, together with Michell's thin-ship value.

\paragraph{Michell's value.} \citeauthor{Michell1898}'s thin-ship
integral for the standard Wigley hull ($B/L=0.10$, $T/L=0.0625$, hull
function separable in $x$ and $z$) was implemented independently in a
short Python script (\texttt{michell\_wigley.py}), not
taken from a published table: closed-form expressions for both
one-dimensional factors of the source-strength transform, a
$\lambda=\cosh t$ substitution to remove the integrable endpoint
singularity of the classical wave-number integral, Simpson quadrature
in $t$ with an explicit tail-convergence check (re-evaluated at a
larger truncation, rejected if the two disagree by more than
$10^{-4}$ relative), and a self-check of both closed-form factors
against direct numerical quadrature at several sample points, run
every time and aborting on failure. $C_w$ is referenced to the same
true (double-sided) wetted surface convention used for SIREN's own
$C_w$. The script depends only on NumPy and never touches SIREN or its
files; its output feeds \texttt{mich.dat}, used throughout this
section and \S\ref{sec:gridconv-farfield}.

\begin{table}[htbp]
\centering
\caption{$C_w$ and $C_t$ ($\times10^3$) against experiment, linear and
non-linear. Computed $C_w$ is $f_\infty$ where Table~\ref{tab:gridconv}/
\ref{tab:gridconv-nl} gives a well-defined apparent order ($p\ge0.5$),
Fine otherwise (marked~$^*$: ``n.m.'' and ``low~$p$'' rows alike,
Richardson extrapolation not trusted in either case); $C_t$ adds
$C_{f,\mathrm{ITTC}}$; err.: $100\,(C_t/C_{t,\mathrm{exp}}-1)$.}
\label{tab:exp}
\begin{tabular}{lr rrr rrr}
\toprule
 & & \multicolumn{3}{c}{linear} & \multicolumn{3}{c}{non-linear} \\
$\Fn$ & $C_{t,\mathrm{exp}}$ & $C_w$ & $C_t$ & err.(\%) & $C_w$ & $C_t$ & err.(\%) \\
\midrule
0.211 & 3.720 & 0.678          & 4.123 & +10.8 & 0.710$^*$ & 4.155 & +11.7 \\
0.250 & 3.940 & 0.995$^*$      & 4.333 & +10.0 & 1.128     & 4.466 & +13.4 \\
0.267 & 4.150 & 0.855$^*$      & 4.153 & +0.1  & 0.887     & 4.186 & +0.9  \\
0.289 & 4.730 & 1.444          & 4.695 & $-$0.7& 1.358$^*$ & 4.609 & $-$2.6\\
0.316 & 5.220 & 1.658          & 4.857 & $-$7.0& 1.802     & 5.001 & $-$4.2\\
0.354 & 5.180 & 1.294$^*$      & 4.429 & $-$14.5&1.453$^*$ & 4.588 & $-$11.4\\
0.408 & 6.350 & 2.436          & 5.493 & $-$13.5&1.791$^*$ & 4.848 & $-$23.7\\
0.450 & 6.710 & 3.354          & 6.360 & $-$5.2 &2.841$^*$ & 5.847 & $-$12.9\\
\bottomrule
\end{tabular}
\end{table}

The two solutions agree with experiment to within the same order of
magnitude everywhere, and neither is uniformly closer: the non-linear
correction improves agreement in the main resistance hump,
$\Fn=0.316$--$0.354$ (error reduced, $-7$/$-15\%$ down to
$-4$/$-11\%$), is close to neutral at $\Fn=0.267$--$0.289$, and is
worse at both ends of the range --- $\Fn=0.211$--$0.250$, where
$C_{w,\mathrm{exp}}$ is itself a small, uncertain difference of $C_t$
and $C_f$, and $\Fn=0.408$--$0.450$, where Table~\ref{tab:gridconv-nl}
already flags the Fine-grid value as not grid-convergent and therefore
not a reliable number to compare in the first place. This is the same
qualitative picture as with the binary wet/dry marking (not shown):
enabling the area-weighted partial-panel correction changes the numbers by a few percent at
every entry but not which regions of the curve the non-linear
correction helps or hurts. Both solutions lie below Michell at most
speeds where it is available; Michell and experiment themselves are
not always on the same side of each other, so their difference is not
by itself an error estimate for either computation. Attitude is fixed in every run
reported here, while the tank model ran free to sink and trim, which
is expected to raise the resistance and is not accounted for above.

\begin{figure}[htbp]
\centering
\includegraphics[width=0.95\textwidth]{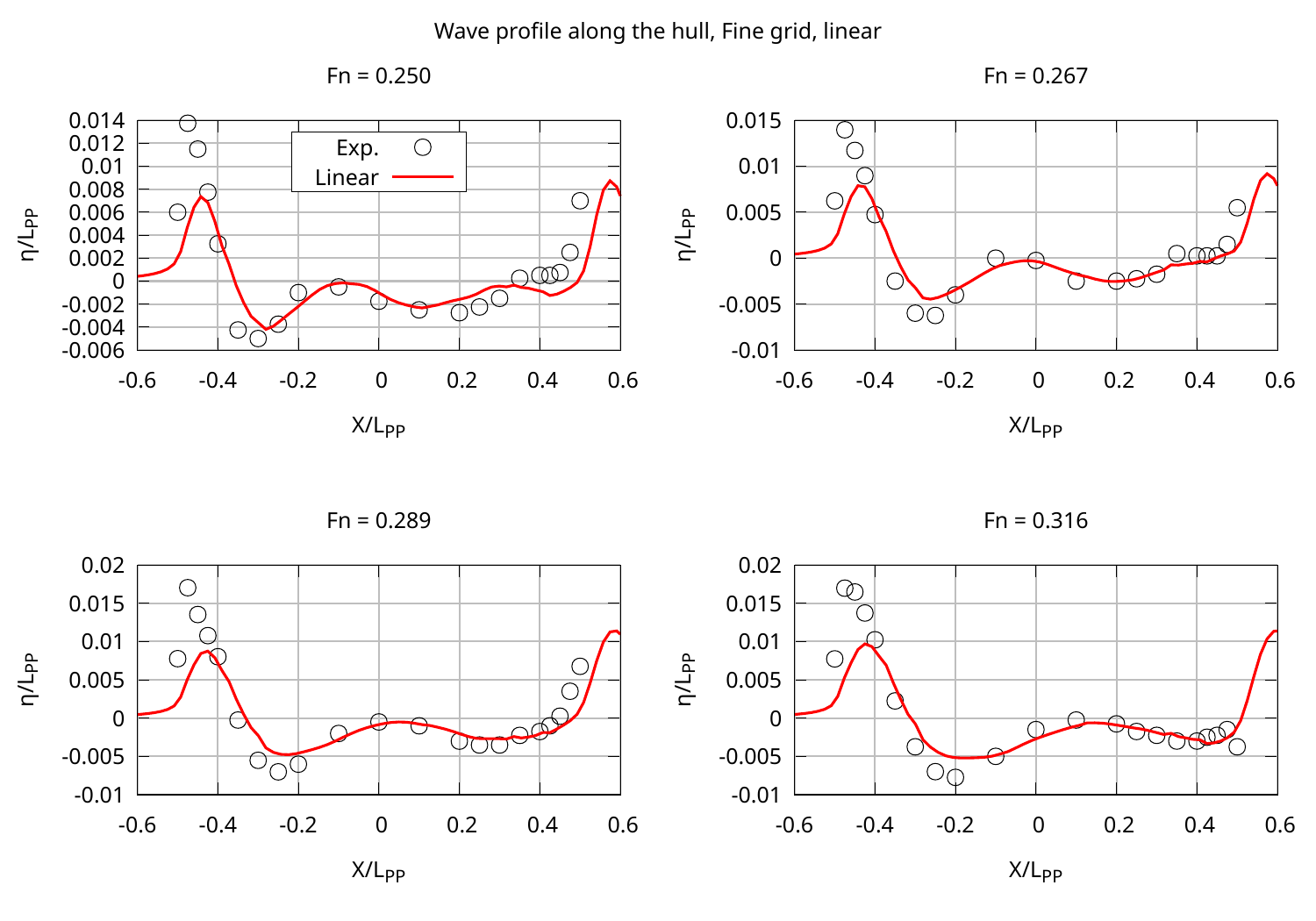}
\caption{Wave elevation along the hull, linear, Fine grid, against the
towing-tank profile, at the four Froude numbers with experimental
data.}
\label{fig:etaL}
\end{figure}

\begin{figure}[htbp]
\centering
\includegraphics[width=0.95\textwidth]{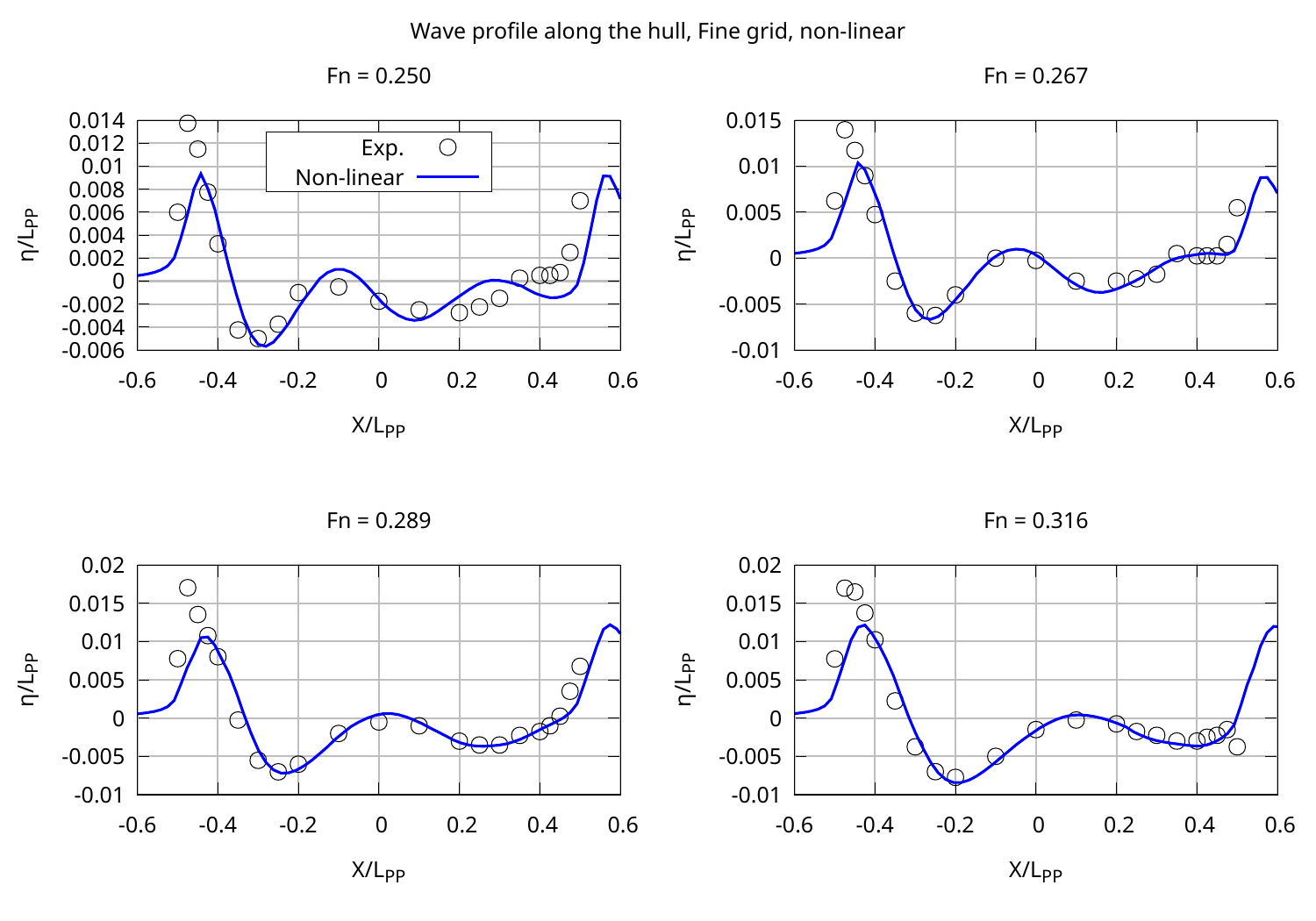}
\caption{As Figure~\ref{fig:etaL}, non-linear solution.}
\label{fig:eta}
\end{figure}

\paragraph{Wave profile along the hull.} Figures~\ref{fig:etaL}
and~\ref{fig:eta} compare the computed wave elevation along the hull,
Fine grid, with the measured Wigley-hull profile of
\citet{Kajitani1983} (Ship Research Institute, Tokyo facility, fixed
condition), linear and non-linear, at
the four Froude numbers for which the experimental profile is
available. The non-linear profile (Figure~\ref{fig:eta}) tracks the
measured trough and stern rise noticeably better than the linear one
(Figure~\ref{fig:etaL}) at every speed shown, consistent with
Table~\ref{tab:exp}'s finding that the non-linear correction improves
agreement in this speed range. Both, however, \emph{underpredict the
bow crest} at every Froude number, by a roughly constant fraction of
its height; the non-linear profile is closer but does not close the
gap. A likely cause is free-surface grid resolution at the bow, where
the wave elevation and its curvature are largest and change fastest
with $x$: this has not been tested (a finer bow-local mesh, or a
fourth overall density, would isolate it) and is noted as the most
likely explanation rather than a demonstrated one.

\subsection{Effect of the nonlinear free-surface iteration}
\label{sec:nonlin-eff}

Earlier configurations of the coupled nonlinear/moving-panel scheme,
tried at various points in this project and not reported individually
here, diverged somewhere in $\Fn=0.211$--$0.450$ on at least one of the
three grids, and where they did not diverge outright, reduced $C_w$
below the linear value everywhere --- the opposite of the physically
expected sign for a wetted surface that follows the bow wave. Both
symptoms traced to the same mechanism: the wet/dry exclusion then in
the linear system (\S\ref{sec:wetsurf}) let the small set of passes
that accumulate the nonlinear correction without resetting it
(\S\ref{sec:nonlin}) compound an error that a single, fresh, fully-relaxed
pass does not show. Removing the exclusion and confining
under-relaxation to exactly those accumulating passes
(\S\ref{sec:nonlin}) removed the divergence (\S\ref{sec:v8}) and
restored the expected sign (Table~\ref{tab:v8sweep}); the resulting
grid convergence and comparison with experiment are given in
Tables~\ref{tab:gridconv-nl} and~\ref{tab:exp} above.

\begin{table}[htbp]
\centering
\caption{Near/far-field consistency, $C_{w,\mathrm{hav}}/C_w$,
area-weighted partial-panel correction enabled, same runs as
Table~\ref{tab:gridconv-nl}
(non-linear) and Table~\ref{tab:gridconv} (linear). $C_{w,\mathrm{hav}}$
does not depend on that correction (\S\ref{sec:wetsurf}: it only
reweights the pressure integral), so this ratio moves only through its
denominator.}
\label{tab:nearfar-nl}
\begin{tabular}{lrrrrrr}
\toprule
 & \multicolumn{3}{c}{linear} & \multicolumn{3}{c}{non-linear} \\
$\Fn$ & Coarse & Medium & Fine & Coarse & Medium & Fine \\
\midrule
0.211 & 1.17 & 1.08 & 1.23 & 1.70 & 1.33 & 1.36 \\
0.250 & 0.97 & 1.17 & 1.14 & 1.20 & 1.24 & 1.20 \\
0.267 & 0.99 & 1.22 & 1.14 & 1.14 & 1.25 & 1.45 \\
0.289 & 1.14 & 1.20 & 1.19 & 1.29 & 1.45 & 1.39 \\
0.316 & 1.08 & 1.22 & 1.22 & 1.05 & 1.13 & 1.17 \\
0.354 & 1.32 & 1.65 & 1.26 & 1.42 & 1.61 & 1.58 \\
0.408 & 1.21 & 1.47 & 1.35 & 1.45 & 1.55 & 1.77 \\
0.450 & 1.35 & 1.55 & 1.71 & 1.24 & 1.55 & 1.50 \\
\bottomrule
\end{tabular}
\end{table}

The ratio is above~1 almost everywhere (the energy-flux estimate above the
pressure integral), the only exceptions being linear Coarse at
$\Fn=0.250$ and $0.267$ (just below unity), unlike the near-unity
agreement reported for an earlier formulation in \S\ref{sec:nearfar};
the two are not directly comparable, since the wet/dry treatment
underlying the pressure integral has changed (\S\ref{sec:wetsurf}).
Establishing a new near-unity baseline for the present formulation is
open.

\subsection{Near-field and far-field consistency}
\label{sec:nearfar}

The pressure-integrated $C_w$ can be checked against three independent
far-field estimates, all computed from the same source strengths rather
than sampled from panel values: the transverse spectral cut of
\S\ref{sec:wavecut} (Eggers/Sharma), a longitudinal spectral
cut at fixed $y=y_c$ following \citet{Newman1963} as codified in the
ITTC longitudinal wave-cut procedure \citep{ITTC750202042021}, and a
direct evaluation of the energy/momentum balance across a transverse
control plane, using the first Havelock formula for the wave
resistance in terms of the free-surface velocity field, which requires
no Fourier transform and hence no separate normalisation constant to
verify, as used for wave-pattern analysis by \citet{Ward1964}.
Table~\ref{tab:nearfar} compares all three against the
pressure-integrated $C_w$ on the Fine grid, linear solution,
from an earlier formulation of the wet/dry
treatment (\S\ref{sec:wetsurf}); the near-unity energy-flux agreement
below is specific to that formulation and is not reproduced by the
present one (\S\ref{sec:nonlin-eff}).

\begin{table}[htbp]
\centering
\caption{Near-field/far-field consistency, Fine grid,
$C_w\times10^3$, linear solution. $C_{w,\mathrm{hav}}$:
exact energy flux; $C_{w,\mathrm{cut}}$:
transverse spectral wave-cut (best-window estimate); ``--'':
cut position falls outside the free-surface domain
(\S\ref{sec:wavecut}, resolution limit) and no value is returned.}
\label{tab:nearfar}
\begin{tabular}{lrrrrr}
\toprule
$\Fn$ & $C_w$ & $C_{w,\mathrm{hav}}$ & hav/press & $C_{w,\mathrm{cut}}$ & cut/press \\
\midrule
0.150 & 0.178 & 0.182 & 1.02 & 0.091 & 0.51 \\
0.180 & 0.348 & 0.235 & 0.67 & 0.204 & 0.58 \\
0.211 & 0.623 & 0.463 & 0.74 & 0.269 & 0.43 \\
0.250 & 0.907 & 0.741 & 0.82 & 0.348 & 0.38 \\
0.267 & 0.830 & 0.659 & 0.79 & 0.355 & 0.43 \\
0.289 & 1.245 & 1.143 & 0.92 & 0.557 & 0.45 \\
0.316 & 1.480 & 1.405 & 0.95 & 0.658 & 0.44 \\
0.354 & 1.233 & 1.274 & 1.03 & 0.482 & 0.39 \\
0.408 & 2.083 & 2.464 & 1.18 & -- & -- \\
0.450 & 2.897 & 3.698 & 1.28 & -- & -- \\
\bottomrule
\end{tabular}
\end{table}

\emph{These values differ slightly (${\sim}1\%$) from the Fine-grid
entries of Table~\ref{tab:gridconv}: the run was repeated after a small
transverse bow refinement was added to the free-surface grid generator,
left at a conservative setting.}

Two results stand out. First, the energy-flux estimate tracks the
pressure-integrated $C_w$ closely over the design-speed range
$\Fn=0.250$--$0.354$ (ratio $0.79$--$1.03$), which is the strongest
evidence available that the near-field solution itself is sound: the
two estimates share nothing but the converged source strengths,
one integrating pressure over the wetted hull, the other integrating
velocity over a plane one to several wavelengths downstream. The
agreement degrades away from this range ($0.67$ at $\Fn=0.180$, $1.28$
at $\Fn=0.450$), in the direction consistent with the free-surface
discretisation error already identified in Table~\ref{tab:gridconv}
($\mathrm{GCI}_{21}>40\%$ at $\Fn=0.267$, $0.354$) rather than with a
defect specific to either method.

Second, the transverse wave-cut underpredicts $C_w$ by a roughly constant
factor (cut/press $=0.38$--$0.58$, no trend with $\Fn$) over the whole
range where it returns a value at all, and returns none above
$\Fn=0.354$ because the required cut position exceeds the free-surface
domain at these longer wavelengths. A constant multiplicative offset
with no dependence on $\Fn$ is the signature of a normalisation
constant, most likely in the kernel of \eqref{eq:rwcut} or in the
factor of two of $H(u)$; it has not been located, and the transverse
wave-cut should be read as qualitative until it is.

The longitudinal wave-cut is not included in Table~\ref{tab:nearfar} because
it does not converge to a single value: the ITTC procedure gives a
result that must be independent of $y_c$, which is the internal check
reported in Table~\ref{tab:lonconsist} rather than a comparison with
$C_w$. Two defects were found and corrected during this study. First,
the wave elevation entering the Fourier transform must be scaled by
$1/\Fn^2$ (it is a length, exactly as $x$ and $y_c$ already were), which
had been omitted; $R_w$ being quadratic in the transform, this alone
had made the raw result smaller than the transverse wave-cut by two to three
orders of magnitude, rather than the comparable order expected of two
methods integrating the same far field. Second, the ITTC procedure's
truncation treatment
--- an analytic tail $\left(c_1\cos x - c_2\sin x\right)/\sqrt{x}$,
fitted by linear regression to the last half of the computed cut and
used to extend the transform numerically some $150$ length units
past the domain edge, in place of the closed-form Fresnel-integral
correction of the reference --- was implemented; this is necessary
because the present free-surface domain gives a cut only about one
ship length long, well short of the reference procedure's own
implicit assumption of a cut long enough for the untreated truncation
error to be negligible.

\begin{table}[htbp]
\centering
\caption{Longitudinal wave-cut internal consistency, Coarse and Fine
grids, linear solution: $C_w\times10^3$ from cuts at $y_c=0.4$ and
$y_c=0.9$, which the method requires to agree. Ratio: larger over
smaller.}
\label{tab:lonconsist}
\begin{tabular}{lrrrrrr}
\toprule
 & \multicolumn{3}{c}{Coarse} & \multicolumn{3}{c}{Fine} \\
$\Fn$ & $y_c=0.4$ & $y_c=0.9$ & ratio & $y_c=0.4$ & $y_c=0.9$ & ratio \\
\midrule
0.211 & 0.167 & 0.061 & 2.75 & 0.244 & 0.131 & 1.86 \\
0.316 & 0.378 & 0.339 & 1.12 & 0.520 & 0.283 & 1.84 \\
0.354 & 0.315 & 0.162 & 1.95 & 0.425 & 0.135 & 3.15 \\
0.450 & 1.711 & 0.476 & 3.60 & 2.254 & 0.560 & 4.02 \\
\bottomrule
\end{tabular}
\end{table}

The truncation treatment is a substantial correction (Table~\ref{tab:lonconsist}:
before it, the Coarse-grid ratio at $\Fn=0.316$ was $1.85$; the
dimensional fix alone does not touch it, since it scales both cuts
equally) but not a complete one, and --- unexpectedly --- it is
\emph{less} effective on the finer grids, where the underlying wave
field is presumably better resolved and carries more genuine
high-wavenumber content than the two-parameter asymptotic model
captures. The longitudinal wave-cut is therefore left as a diagnostic
under development rather than a third independent resistance estimate:
useful for flagging gross errors (the dimensional bug it caught would
have been invisible to the energy-flux and transverse wave-cut
estimates alone, since both already used correctly-scaled quantities),
not yet reliable enough to quote a value from.

Taken together, the three checks separate cleanly by trust level: the
energy-flux estimate corroborates the near-field solution to within
the grid's own discretisation error and is the recommended far-field
cross-check; the transverse wave-cut is qualitatively consistent (right
order of magnitude, no sign errors, smooth trend with $\Fn$) but
carries an unresolved constant offset; the longitudinal wave-cut is
informative for catching implementation errors but not yet
self-consistent enough to report.

\subsection{Grid convergence of the far-field estimates}
\label{sec:gridconv-farfield}

The same three-grid Richardson/GCI procedure of \S\ref{sec:gridconv},
applied there to the pressure-integrated $C_w$, was repeated for all
three far-field estimates of \S\ref{sec:nearfar}--\S\ref{sec:wavecut}
(energy-flux, transverse wave-cut, longitudinal wave-cut at
$y_c=0.4$), at every Froude number, linear and non-linear: $144$
grid/method/Froude values in total, $48$ three-grid sequences. The
pressure integral is monotonic in grid density at six of the eight
Froude numbers, both linear and non-linear, with a well-defined
apparent order at all six linear and three of the six non-linear
sequences (Tables~\ref{tab:gridconv}, \ref{tab:gridconv-nl}: $9$ of
$16$ (Froude, formulation) pairs with a usable $\mathrm{GCI}$). The
far-field sequences are monotonic far more rarely: $7$ of the $48$
(three for the energy-flux estimate, linear; two for the energy-flux
estimate, non-linear; two for the transverse wave-cut, non-linear; the
longitudinal wave-cut, either formulation, and the transverse
wave-cut, linear, are monotonic nowhere in this Froude range), spread
over only $6$ of the $16$ (Froude, formulation) pairs, and one further
energy-flux sequence (linear, $\Fn=0.450$) is monotonic but with
$p=0.42$, an apparent order too low to trust
($\mathrm{GCI}_{21}=63\%$, asymptotic ratio $82$, excluded from
Table~\ref{tab:gridconv-farfield} for the same reason the low-$p$ rows
of Table~\ref{tab:gridconv-nl} are). Where a far-field sequence
\emph{is} genuinely convergent, its Richardson value agrees with the
towing-tank $C_{w,\mathrm{exp}}$ to within a few tens of percent, the
same order of agreement as the pressure-integrated estimate
(Table~\ref{tab:exp}) --- but at well under half as many
(Froude, formulation) pairs. The far-field estimates remain useful
cross-checks of the near-field solution at the single grid where they
are evaluated (\S\ref{sec:nearfar}), but --- with the present
free-surface discretisation --- only the pressure integral is
grid-convergent enough, across the whole Froude range, to support an
asymptotic value with a defined uncertainty band.

\begin{table}[htbp]
\centering
\caption{Far-field $C_w\times10^3$ grid convergence, the $7$ (of $48$)
grid/method/Froude sequences found monotonic with a well-defined
apparent order ($p\ge0.5$); layout as Table~\ref{tab:gridconv}. All
other sequences are non-monotonic (or, one case, monotonic with
$p<0.5$) in this Froude range and are not reported here.}
\label{tab:gridconv-farfield}
\begin{tabular}{llrrrrr}
\toprule
formulation & method & $\Fn$ & $p$ & $f_\infty$ & $\mathrm{GCI}_{21}$ (\%) & $C_{w,\mathrm{exp}}$ \\
\midrule
linear     & energy-flux           & 0.289 & 7.06 & 1.704 & 0.0  & 1.471 \\
linear     & energy-flux           & 0.316 & 16.00& 2.017 & 0.0  & 2.008 \\
non-linear & energy-flux           & 0.316 & 1.53 & 2.227 & 6.9  & 2.008 \\
non-linear & energy-flux           & 0.354 & 6.48 & 2.294 & 0.1  & 2.027 \\
non-linear & transverse wave-cut   & 0.211 & 3.06 & 0.381 & 13.1 & 0.273 \\
non-linear & transverse wave-cut   & 0.250 & 6.70 & 0.617 & 0.2  & 0.598 \\
non-linear & transverse wave-cut   & 0.354 & 2.37 & 0.881 & 1.8  & 2.027 \\
\bottomrule
\end{tabular}
\end{table}

The non-linear transverse-cut row at $\Fn=0.354$ stands out: $f_\infty$
is grid-convergent in the Roache/Celik sense, yet differs from
$C_{w,\mathrm{exp}}$ by $57\%$ --- consistent with the cut's own
unresolved constant offset (\S\ref{sec:wavecut}) rather than a
contradiction of it: a sequence can be genuinely convergent to the
wrong constant.

%=======================================================================
\section{Discussion}
%=======================================================================

The base flow is the point of departure from the classical scheme. The
double-body flow is exact for a flat lid and a hull ending at $z=0$, but
as soon as the wetted hull extends above $z=0$ its image overlaps the
hull itself; a base flow defined on the actual geometry does not have
this problem. Imposing impermeability on the whole hull, wet and dry,
rather than exchanging panels in and out of the linear system as the
wave moves, removed a second, independent source of instability
(\S\ref{sec:wetsurf}, \S\ref{sec:nonlin-eff}) and, as a side effect,
the pressure field at freshly-wetted bow panels, which a panel excluded
one pass earlier had never actually been solved for.

What is now established, on the Wigley hull at $\Fn=0.211$--$0.450$
(Fine at $0.150$, $0.180$ too): the non-linear iteration is stable on
all three grids at every Froude number but one; the grid sequence is
either tightly converged or cleanly monotonic with a well-defined GCI
at most speeds, linear and non-linear; and the non-linear correction
has the physically expected sign (raises $C_w$) and improves agreement
with experiment in the main resistance hump, though not everywhere
(\S\ref{sec:linref}). What remains open: neither iteration reaches its
input tolerance in the number of passes used here; the $\Fn=0.408$,
$0.450$ non-linear results are grid-\emph{inconsistent}, not merely
grid-coarse (Table~\ref{tab:gridconv-nl}); the near/far-field ratio
moved away from unity under the present wet/dry treatment
(\S\ref{sec:nonlin-eff}) and has not been re-established; catamaran and
transom-stern cases are untested with the present formulation; and the
free-surface panels satisfy the exact condition where they sit, not on
the true (sloped) surface, since they are moved rigidly in the
vertical.

%=======================================================================
\section{Conclusions}
%=======================================================================

A Rankine-source solver has been developed in which the base flow is a
rigid-lid Neumann problem on the actual hull and free-surface geometry,
with no image system and no desingularisation; impermeability is
imposed on the whole hull, wet and dry, at every pass, with the wet/dry
marking used only to filter the force integral; and under-relaxation is
confined to the passes where the non-linear correction genuinely
accumulates. On a Wigley hull at $\Fn=0.150$--$0.450$, three free-surface
densities, the non-linear iteration is stable everywhere but the lowest
Froude number on the finest grid; grid convergence and comparison with
Michell and experiment are reported for both the linear and non-linear
solutions (\S\ref{sec:gridconv}--\S\ref{sec:linref}); and the non-linear
correction raises $C_w$, as expected, and improves agreement with
experiment in the main resistance hump. Open points are listed in the
Discussion above.

These results are preliminary. They cover a single hull, fixed
attitude, at model scale, and every open point listed in the
Discussion --- the iterative tolerance not reached within the passes
used here, grid-inconsistency at the two highest non-linear Froude
numbers, the far-field estimates' patchy grid convergence
(\S\ref{sec:gridconv-farfield}), the near/far-field baseline still to
be re-established --- awaits further verification before any of it is
relied upon beyond this paper. Two further extensions are immediate
and, in the strict sense of requiring new code, unnecessary: free
sinkage and trim, and the twin-hull (SWATH/catamaran) and transom-stern
geometries are already implemented (\S\ref{sec:sinkagetrim},
\S\ref{sec:transom}, \S\ref{sec:capabilities}) but not exercised by any
result reported above, and both are needed before the present findings
can be said to extend beyond a fixed-attitude monohull.

%=======================================================================
\section*{Acknowledgements}
%=======================================================================

Claude (Anthropic), an AI assistant, was used under the author's
direction and review as a technical collaborator throughout this
work for the final check and optiomization of code implementation.

%=======================================================================

\end{document}